\documentclass[runningheads]{llncs}

\usepackage{booktabs}   
\usepackage{caption}    
\usepackage{multirow}   
\usepackage{makecell} 
\usepackage{tabularx} 
\usepackage{subcaption}
\usepackage[T1]{fontenc}
\usepackage{amsmath,amsfonts}

\usepackage{algorithmic}
\usepackage{algorithm}
\usepackage{array}
\usepackage[caption=false,font=normalsize,labelfont=sf,textfont=sf]{subfig}
\usepackage{textcomp}
\usepackage{stfloats}
\usepackage{url}
\usepackage{verbatim}
\usepackage{graphicx}
\usepackage{cite}
\usepackage{caption}
\usepackage{subcaption}

\usepackage{pgfplots}
\usepgfplotslibrary{groupplots}
\pgfplotsset{compat=1.18}

\usepackage{tabularx}
\usepackage{subcaption}
\usepackage{booktabs}
\usepackage{array}

\usepackage{listings}
\lstdefinelanguage{JavaScript}{
  keywords={break, case, catch, continue, debugger, default, delete, do, else, finally, for, function, if, in, instanceof, new, return, switch, this, throw, try, typeof, var, void, while, with},
  morecomment=[l]{//},
  morecomment=[s]{/*}{*/},
  morestring=[b]',
  morestring=[b]",
  sensitive=true
}
\usepackage{xcolor}
\definecolor{codegreen}{rgb}{0,0.6,0}
\definecolor{codegray}{rgb}{0.5,0.5,0.5}
\definecolor{codepurple}{rgb}{0.58,0,0.82}
\definecolor{backcolour}{rgb}{0.95,0.95,0.92}

\usepackage{caption}
\lstdefinestyle{mystyle}{
    backgroundcolor=\color{backcolour},   
    commentstyle=\color{codegreen},
    keywordstyle=\color{magenta},
    numberstyle=\tiny\color{codegray},
    stringstyle=\color{codepurple},
    basicstyle=\ttfamily\footnotesize,
    breakatwhitespace=false,         
    breaklines=true,                 
    captionpos=b,                    
    keepspaces=true,                 
    numbers=left,                    
    numbersep=5pt,                  
    showspaces=false,                
    showstringspaces=false,
    showtabs=false,                  
    tabsize=2
}

\colorlet{punct}{red!60!black}
\definecolor{background}{HTML}{EEEEEE}
\definecolor{delim}{RGB}{20,105,176}
\colorlet{numb}{magenta!60!black}

\lstdefinelanguage{json}{
    numbers=left,
    numberstyle=\scriptsize,
    stepnumber=1,
    numbersep=8pt,
    breaklines=true,
    backgroundcolor=\color{background},
    literate=
     *{0}{{{\color{numb}0}}}{1}
      {1}{{{\color{numb}1}}}{1}
      {2}{{{\color{numb}2}}}{1}
      {3}{{{\color{numb}3}}}{1}
      {4}{{{\color{numb}4}}}{1}
      {5}{{{\color{numb}5}}}{1}
      {6}{{{\color{numb}6}}}{1}
      {7}{{{\color{numb}7}}}{1}
      {8}{{{\color{numb}8}}}{1}
      {9}{{{\color{numb}9}}}{1}
      {:}{{{\color{punct}{:}}}}{1}
      {,}{{{\color{punct}{,}}}}{1}
      {\{}{{{\color{delim}{\{}}}}{1}
      {\}}{{{\color{delim}{\}}}}}{1}
      {[}{{{\color{delim}{[}}}}{1}
      {]}{{{\color{delim}{]}}}}{1},
}

\newcolumntype{C}{>{\centering\arraybackslash}X} 
\newcolumntype{L}{>{\raggedright\arraybackslash}X} 

\begin{document}

\title{Trade-Size-Aware Dynamic Fees for Impermanent Loss Mitigation in AMMs}

\author{Anton Ledrov\inst{1} \and
Ignat Melnikov\inst{2} \and
Irina Lebedeva\inst{2} \and
Dmitrii Umnov\inst{3} \and \\
George Ovchinnikov\inst{2} \and
Yury Yanovich\inst{2}}
\titlerunning{Trade-Size-Aware Dynamic Fees}
\authorrunning{A. Ledrov et al.}
%

\institute{Moscow Institute of Physics and Technology, Moscow, Russia  \and
Skolkovo Institute of Science and Technology, Moscow, Russia \and
HSE University, Moscow, Russia}

\maketitle

\begin{abstract}
Automated Market Makers enable decentralized trading but systematically expose liquidity providers to impermanent loss through arbitrage-driven rebalancing. While dynamic fee mechanisms offer a promising mitigation strategy, existing approaches remain largely reactive, adjusting costs based on historical signals rather than explicitly linking them to the structural risk imposed by individual trades. To address this limitation, we propose a novel fee formation framework built on three core innovations. First, we introduce a coupled market maker architecture in which fee dynamics are governed by a secondary invariant, allowing liquidity state and transaction costs to evolve jointly. Second, we develop an impermanent-loss trimming fee model that adaptively increases transaction costs for trades exceeding the liquidity providers' profitable region, effectively offsetting losses from large arbitrage executions while preserving baseline fees for smaller transactions. Third, we establish a unified evaluation methodology using performance profiles to systematically compare fee algorithms across diverse market conditions.  By extending a heterogeneous trader model to derive optimal arbitrage strategies under state-dependent fees, we conduct extensive simulations on historical data spanning four distinct market regimes and three token pair categories. Our results demonstrate that the proposed fee enhancements improve liquidity provider yields by 6--24\% in volatile markets and up to 119\% in calm regimes, while maintaining uninformed user participation and reducing informed arbitrage profitability by 3--10\%. Performance profile analysis confirms that ILT-enhanced algorithms dominate baseline counterparts across 60--75\% of test scenarios. These findings indicate that structurally grounded, trade-aware fee mechanisms can significantly enhance the economic sustainability of decentralized liquidity provision.

\keywords{AMM \and Blockchain \and DEX \and Dynamic Fee \and Impermanent Loss \and Performance Profiles}
\end{abstract}

\section{Introduction}
\label{sec:introduction}
Automated Market Makers~(AMMs) have become a foundational mechanism for Decentralized Exchanges (DEXs), enabling continuous liquidity provision without relying on order books used by Centralized Exchanges (CEXs)~\cite{pourpouneh2020automated}. By replacing discrete matching with invariant-based pricing, AMMs offer a scalable and composable alternative to traditional exchange design. However, in this model liquidity providers~(LPs) are systematically exposed to adverse selection and value extraction through arbitrage~\cite{wang2022cyclic}. A central manifestation of this issue is impermanent loss~(IL), which arises when the composition of assets in a liquidity pool adjusts to external price movements~\cite{aigner2021uniswap}. Arbitrageurs, who have access to global price information, rebalance the pool by trading against stale on-chain prices, effectively transferring value from LPs to themselves~\cite{boonpeam2021arbitrage}. As a result, LP profitability depends critically on the ability of the protocol to capture sufficient fees from trading activity to offset these losses.

Traditional AMMs rely on fixed transaction fees, which are simple to implement but fundamentally misaligned with the dynamic nature of market risk. A single fee level cannot simultaneously compensate LPs during volatile market periods and remain attractive to users during stable ones~\cite{hasbrouck2022need}. In response, dynamic fee mechanisms have moved from optional extensions to core components of modern DEXs. For example, Uniswap v4 introduces a hook-based architecture that allows developers to attach custom fee logic to predefined lifecycle callbacks, enabling direct implementation of dynamic policies within the swap execution pipeline~\cite{uniswapv4}. 

Prior work ~\cite{lebedeva2025dynamic_fee} explored adaptive fee mechanisms that reduce IL and improve LP outcomes through directional and activity-based fee adjustments. However, such mechanisms remain largely reactive: they depend on historical signals and do not explicitly encode the structural relationship between trade characteristics and the risk imposed on liquidity providers.


We propose a new fee formation approach based on three core ideas. First, we introduce an~\emph{AMM-in-AMM} architecture, in which the fee mechanism is itself governed by an invariant structure: fee dynamics arise endogenously from a secondary market maker coupled to the primary pool. Second, we formalize the notion of an~\emph{Impermanent Gain~(IG)} region in the space of admissible trades~\cite{Vlasov2025Impact, melnikov2026impermanent_gain} -- the set of swaps that impose limited risk on LPs and may therefore be executed at baseline cost. Third, we develop an~\emph{Impermanent-Loss Trimming~(ILT)} fee model that scales transaction costs with the market impact of a trade, selectively offsetting IL for information-driven swaps that are larger than the IG threshold while preserving baseline costs for smaller transactions. The ILT layer is non-intrusive and integrates as a wrapper over existing dynamic fee algorithms.  

Extending the heterogeneous trader model of~\cite{lebedeva2025dynamic_fee}, we analyze how endogenous, state-dependent fees reshape the structure of arbitrage opportunities and improve the alignment between trading costs and the economic risk borne by LPs.

The main contributions of this work are as follows:
\begin{itemize}
    \item We propose a novel endogenous fee formation mechanism based on a coupled \emph{AMM-in-AMM} architecture, enabling transaction costs and liquidity state to evolve jointly.
    \item We formalize the Impermanent Gain region and derive an \emph{Impermanent-Loss Trimming} fee model that scales costs with trade size, and demonstrate that it integrates with and enhances existing dynamic fee algorithms.
    \item We conduct a comprehensive empirical evaluation across four market regimes and three token pair categories, showing that ILT-enhanced algorithms improve LP yields by 6--24\% in volatile markets and up to 119\% in calm regimes, while preserving uninformed user participation ($\approx -57$ bps yield unchanged) and reducing informed arbitrage profitability by 3--10\%. Performance profile analysis confirms ILT dominance across 60–75\% of test scenarios.
\end{itemize}

Our results suggest that moving from reactive fee adjustment to structurally grounded fee formation can enhance the efficiency and robustness of AMM-based markets.

\section{Related Work}
\label{sec:related_work}
AMMs constitute the dominant design for DEXs, replacing order books with invariant-based pricing functions \cite{pourpouneh2020automated, Angeris2020}. Trades move the pool state along a predefined curve, inducing slippage and creating arbitrage opportunities relative to external markets. The central risk faced by LPs in this setting is IL, defined as the difference between the value of the LP's position and the value of holding the same assets externally~\cite{aigner2021uniswap, tangri2023generalizing}. IL is closely linked to arbitrage activity: arbitrageurs rebalance the pool against price discrepancies between DEXs and CEXs~\cite{wang2022cyclic, boonpeam2021arbitrage}, while trade fees constitute the primary source of LP revenue, producing a fundamental trade-off between risk and compensation. Recent work refines this picture by introducing IG region~\cite{melnikov2026impermanent_gain}: when price deviations are sufficiently small, fee income exceeds rebalancing losses, and arbitrage activity yields a net positive outcome for LPs.

Closer to our work, in~\cite{lebedeva2025dynamic_fee} a heterogeneous trader model has been proposed. Participants are divided into Informed Users~(IU) -- who act as arbitrageurs and optimize trades against external price information -- and Uninformed Users~(UU) -- who trade for exogenous reasons modulated by a psychological loss factor that filters execution by perceived quality. Within this framework, the authors introduce several adaptive fee mechanisms (block-adaptive, deal-adaptive, oracle-based) that mitigate IL by adjusting fees in response to recent trading direction and activity. While these approaches demonstrate empirical improvements over fixed fees, they remain reactive in nature and do not explicitly link fee levels to the structural risk imposed by individual trades -- in particular, to trade size and its nonlinear effect on adverse selection.

Alternative AMM designs address the LP-arbitrage trade-off through different mechanisms. The auction-managed AMM~(am-AMM) assigns fee control to a manager via on-chain auction mechanism, enabling partial internalization of arbitrage opportunities and improving equilibrium liquidity relative to fixed-fee pools at the cost of increased strategic complexity~\cite{Adams2024amAMMAA}. Broader work on adaptive fee mechanisms has shown that conditioning fees on real-time volatility and order flow imbalance enhances capital efficiency and reduces IL~\cite{Milionis2023AutomatedMM}. In another related direction, the authors examine the interaction between slippage, liquidity depth, and arbitrage incentives, proposing a mechanism that adjusts effective pool depth and showing that increased liquidity stabilizes prices but must be calibrated carefully to preserve arbitrage activity~\cite{wen2026balance}.

From a theoretical perspective, fee design has been studied as a stochastic control problem in which optimal fees balance arbitrage deterrence and trading volume~\cite{dynAMM}, and as a game-theoretic problem in which competing pools converge to equilibrium fee levels shaped by trader-LP interaction~\cite{Fritsch_2021}. These results highlight that fee setting is inherently an endogenous process shaped by participant behavior.

A recent systematic review of IL in AMMs reports that most existing studies focus on constant product market makers (CPMMs), reflecting their dominant role in practice, while quantitative evaluation and empirical validation remain comparatively limited~\cite{delmonte2025impermanent}. In line with this gap, Urusov et al.~\cite{urusov2025backtesting} develop a backtesting framework for Uniswap V3 that models liquidity distribution, pool state transitions, and fee accumulation dynamics with sub-1\% error against historical data, demonstrating the value of simulation-based evaluation for LP performance.

Beyond AMMs, dynamic fee design is part of a broader pattern in which DeFi protocols expose critical parameters whose performance depends sensitively on strategic user behavior. This phenomenon is observed not only in AMM fee design~\cite{Fritsch_2021, dynAMM}, but also in lending protocols, where collateral factors and interest rates are tuned with respect to default risk and system stability~\cite{lending1, lending2}. Parameter selection thus becomes a mechanism-design problem under strategic interaction, rather than a purely technical calibration task.

\section{System Model}
\label{sec:model}

\subsection{DEX and Liquidity Pool Dynamics}

Motivated by the dominant role of CPMMs in both practice and the IL literature~\cite{dynAMM}, we adopt a Uniswap V2-style CPMM as the analytical setting throughout this work, following the formalism of~\cite{Vlasov2025Impact}. Let $(x, y)$ denote the reserves of two assets in the liquidity pool. In Uniswap V2, the system is governed by the constant product invariant:
$
x y = K^2.
$

The instantaneous exchange rate is determined by the marginal price:
\begin{equation}
p(x, K) = -\frac{dy}{dx} = \frac{y}{x}.
\end{equation}

In Uniswap V2, the trading fee is applied only to the input asset and remains in the pool. Let $\phi$ denote the fee rate and $\gamma = 1 - \phi$. For a trade of size $\Delta x$, only $\gamma \Delta x$ participates in price formation, while $(1-\gamma)\Delta x$ is retained in the reserves.

The exchanged amount $\Delta y$ is determined from the conservative relation:
\begin{equation}\label{eq:invarian_save}
(x_0 + \gamma \Delta x)(y_0 - \Delta y) = x_0 y_0.
\end{equation}

Importantly, this equation describes only the swap computation. After the trade, the full amount $\Delta x$ is added to the pool:
\begin{equation*}
x_1 = x_0 + \Delta x, \quad y_1 = y_0 - \Delta y,
\end{equation*}
so that the invariant increases $K_1 > K_0$.

To quantify the effect on liquidity providers, we compare depositing assets into the AMM with holding them externally. Following \cite{Vlasov2025Impact}, the corresponding values are:
\begin{equation*}
V_{\text{deposit}} = p(x_0 + \Delta x, K_1)\cdot (x_0 + \Delta x) + (y_0 - \Delta y),
\end{equation*}
\begin{equation*}
V_{\text{hold}} = p(x_0 + \Delta x, K_1)\cdot x_0 + y_0.
\end{equation*}

The absolute impermanent loss is defined as:
$
IL = p(x_0 + \Delta x, K_1)\cdot \Delta x - \Delta y.
$
Negative values correspond to impermanent loss, while positive values correspond to impermanent gain. Moreover, the exact expression for IL in Uniswap V2 for $X \rightarrow Y$ swap is derived in~\cite{Vlasov2025Impact}:
\begin{equation}\label{eq:il}
IL = y_0 \cdot \frac{\alpha}{1 + \alpha} \cdot \frac{1 - \gamma - \gamma \alpha}{1 + \gamma \alpha}
\end{equation}
where $\alpha = \frac{\Delta x}{x}$. Strict bounds for positive IL (referred to as Impermanent Gain) were also established:
\begin{equation}\label{eq:ig_frame}
\alpha_{\max} = \frac{1 - \gamma}{\gamma}.
\end{equation}
\subsection{Trader Behavior Model}

In the proposed framework, traders interacting with the DEX are divided into two behavioral groups: informed users and uninformed users.

\subsubsection{Informed Users (IU).}
IU are modeled as agents who actively exploit price discrepancies between the DEX and an external reference market (CEX). Their objective is to maximize the change in portfolio value evaluated at external prices.

We define the portfolio value as
$
P(a,b) = a p_A + b p_B,
$
where $p_A$ and $p_B$ denote prices on the external market. The change in capital induced by a trade is given by
\begin{equation}
\delta P = P(\Delta x, \Delta y) - P(f(x,y,\Delta x,\Delta y)) - \mathrm{fee}_\text{tx}.
\end{equation}

The IU chooses trade sizes $(\Delta x, \Delta y)$ as a solution of
$
\max_{\Delta x, \Delta y} \ \delta P,
$
and executes the trade only if the resulting value is positive. This mechanism captures arbitrage behavior and ensures alignment between DEX and external prices~\cite{lebedeva2025dynamic_fee}.

To make this problem explicit, consider a swap $\Delta x \to \Delta y$ under a constant product AMM with fee parameter $\gamma$. The trade is constrained by the invariant~\ref{eq:invarian_save}. For $\alpha = \frac{\Delta x}{x}$, we can express the output amount as
$
\Delta y = y \cdot \frac{\alpha \gamma}{1 + \alpha \gamma}.
$

Substituting into the objective function, we obtain a one-dimensional optimization problem:
\begin{equation}
\Delta P(\alpha) = -p_A x \alpha + p_B y \cdot \frac{\alpha \gamma}{1 + \alpha \gamma}.
\end{equation}

The optimal swap size is determined from the first-order condition
which in the case of a fixed fee gives ($\gamma = \mathrm{const}$):
\begin{equation}
\alpha^* = \frac{1}{\gamma} \left( \sqrt{\frac{p_B y \gamma}{p_A x}} - 1 \right).
\end{equation}


\subsubsection{Uninformed Users (UU).}

Uninformed users are modeled as agents who interact with the automated market maker (AMM) for exogenous, non-arbitrage reasons. Their trading behavior is described as a stochastic process with a behavioral acceptance mechanism that depends on execution quality.

UUs arrive at each discrete time step with probability $p_\text{UU}$. Conditional on arrival, the trade direction is drawn from a symmetric Bernoulli distribution (reflecting the absence of directional information) and the size as a fraction $q \sim \mathcal{N}(\mu_\text{UU}, \sigma^2_\text{UU})$ of the pool reserves, restricted to a bounded interval $q \in (0, q_\text{max})$ to ensure feasibility. The UU then evaluates execution quality through a psychological loss factor defined as
\begin{equation}
r = \frac{\delta P(\Delta x, \Delta y)}{\delta P(|\Delta x|, |\Delta y|)},
\end{equation}
which captures the deviation between expected and realized outcomes after fees and slippage. Here, $\delta P(\cdot)$ represents the change in portfolio value induced by the transaction. Negative values of $r$ correspond to perceived losses due to unfavorable execution.

Trade execution for UU is probabilistic:
\begin{equation}
P_{\text{exec}} =
\begin{cases}
1, & r > 0, \\
\exp(-|r|), & r \leq 0,
\end{cases}
\end{equation}
so that favorable trades are always executed while the likelihood of execution decreases exponentially with perceived losses. The result is a stochastic trade flow filtered by execution quality.

\section{Dynamic Fee Mechanisms}
\label{sec:fee_algorithms}
We evaluate a spectrum of fee mechanisms, separating established reference algorithms from the novel contributions proposed in this work. This taxonomy allows us to isolate the incremental benefits of our new mechanisms while maintaining compatibility with prior adaptive fee designs.

\subsection{Baseline and Prior Adaptive Fees}
These algorithms serve as reference points and are either standard in practice or adapted from prior work on dynamic fee scheduling~\cite{lebedeva2025dynamic_fee}.

\paragraph{Fixed Fee~(FX).}
The baseline model assumes a constant fee applied uniformly to all trades, regardless of direction or market condition:
$
f_{A \to B} = f_{B \to A} = f^{\mathrm{fx}}.
$
Although simple and widely used, this approach does not account for market dynamics and remains vulnerable to arbitrage-induced losses for liquidity providers.

\paragraph{Block-Adaptive Fee~(BA).}
This mechanism exploits temporal correlation in arbitrage activity. At each block, directional fees are updated based on observed price changes:
\begin{itemize}
    \item If price movement indicates arbitrage pressure in the $A \to B$ direction:
    \begin{equation}
    f_{A \to B} \leftarrow f_{A \to B} + \Delta f, \quad
    f_{B \to A} \leftarrow f_{B \to A} - \Delta f.
    \end{equation}
    \item The update is symmetric for the opposite direction.
    \item If no significant price change is detected, fees remain unchanged.
\end{itemize}
This approach increases costs for informed traders while preserving relatively low fees for opposite-direction trades.

\paragraph{Deal-Adaptive Fee~(DA).}
The deal-adaptive algorithm updates fees after each transaction, making it highly responsive to local order flow. After a trade:
\begin{itemize}
    \item If the last trade was $A \to B$:
    \begin{equation}
    f_{A \to B} \leftarrow f_{A \to B} + \delta f, \quad
    f_{B \to A} \leftarrow f_{B \to A} - \delta f.
    \end{equation}
\end{itemize}
The average fee level is constrained:
$
\frac{f_{A \to B} + f_{B \to A}}{2} \leq f^{\mathrm{max}}.
$
This gradually penalizes persistent directional flow (typically associated with arbitrage) while incentivizing counter-flow trades.

\paragraph{Oracle-Based Fee~(OB).}
The oracle-based mechanism represents an idealized benchmark assuming access to external ``fair'' prices. Fees are assigned asymmetrically depending on the arbitrage direction:
$
f_{\mathrm{arb}} = f^{\mathrm{high}}, \quad
f_{\mathrm{non\text{-}arb}} = f^{\mathrm{low}}.
$
This eliminates information asymmetry and provides an upper bound on achievable performance, though it is not directly implementable in practice due to the lack of reliable on-chain external price feeds.

\subsection{Proposed AMM-Based Fee (AB)}
\label{subsec:amm_fee}
We introduce an~\emph{AMM-in-AMM} architecture in which directional transaction fees $(f_{A \to B}, f_{B \to A})$ are treated as the state variables of a secondary market maker governed by a constant-sum invariant $f_{A \to B} + f_{B \to A} = K_{\mathrm{fee}}$, where $K_{\mathrm{fee}}$ is the protocol's target aggregate fee level. Rather than prescribing fee updates through external heuristics, we model fee dynamics as an endogenous process driven by the primary pool's price movements. Each incoming trade triggers a deterministic state transition in the fee pool: when internal price dynamics indicate sustained directional flow, the corresponding fee increases while the opposite fee decreases symmetrically, preserving the invariant. This creates a coupled dual-AMM system where the primary liquidity state $(x,y)$ and the secondary fee state co-evolve, establishing a closed-loop, self-regulating mechanism.

Formally, let $\delta$ denote the observed internal price change within a block. The fee update rule is:
$
f_{A \to B} \leftarrow f_{A \to B} + \beta \delta, \quad
f_{B \to A} \leftarrow f_{B \to A} - \beta \delta,
$
where $\beta > 0$ controls sensitivity. The constant-sum constraint ensures that $f_{A \to B} + f_{B \to A} = K_{\mathrm{fee}}$ holds at all times, effectively treating the fee pair as reserves of a secondary invariant structure.

Beyond bounded directional fee adjustment, the invariant-based formulation provides three structural properties. First, aggregate fees remain stable at $K_\text{fee}$ by construction, eliminating fee drift independent of sensitivity calibration. Second, directional symmetry is enforced architecturally -- any increase in one fee is exactly compensated by the opposite decrease -- rather than imposed as a separate update rule. Third, the parameter $K_\text{fee}$ provides a single, interpretable control parameter for the protocol designer to set the aggregate fee level, decoupled from the sensitivity parameter $\beta$ that controls how rapidly fees respond to flow. Together, these properties guarantee bounded fee dynamics while aligning transaction costs with endogenous market signals.


\subsection{Impermanent-Loss Trimming (ILT) Modification}
\label{subsec:ilt_fee}
We introduce a novel trade-size-aware modification that can be applied as a wrapper to any base fee algorithm. The key insight is that impermanent loss becomes positive (i.e., Impermanent Gain) only when the relative trade size $\alpha = \Delta x / x$ satisfies $\alpha \leq \alpha_{\max} = (1-\gamma)/\gamma$~\eqref{eq:ig_frame}. For larger trades, the LP incurs a net loss.

The ILT mechanism implements a piecewise fee function:
\begin{equation}\label{eq:ilt_piecewise}
\gamma(\alpha) = 
\begin{cases}
\gamma_{\mathrm{fix}}, & \alpha \leq \alpha_\text{max} \\
\gamma_\text{il}(\alpha), & \alpha > \alpha_\text{max}
\end{cases}
=
\begin{cases}
\gamma_{\mathrm{fix}}, & \alpha \leq \alpha_\text{max} \\
\frac{1}{1+\alpha}, & \alpha > \alpha_\text{max}
\end{cases}
\end{equation}
where $\gamma_{\mathrm{fix}}$ is the baseline fee coefficient determined by the underlying algorithm (FX, BA, DA, AB, or OB), and $\gamma_\text{il}(\alpha)$ is chosen to enforce $\mathrm{IL}(\alpha, \gamma_\text{il}) = 0$. Substituting the IL expression~\eqref{eq:il} and solving $\mathrm{IL}=0$ for $\gamma$ yields the closed-form trimming function $\gamma_\text{il}(\alpha) = \frac{1}{1+\alpha}$.

\begin{figure}[!htbp]
    \centering
    \includegraphics[width=0.8\linewidth]{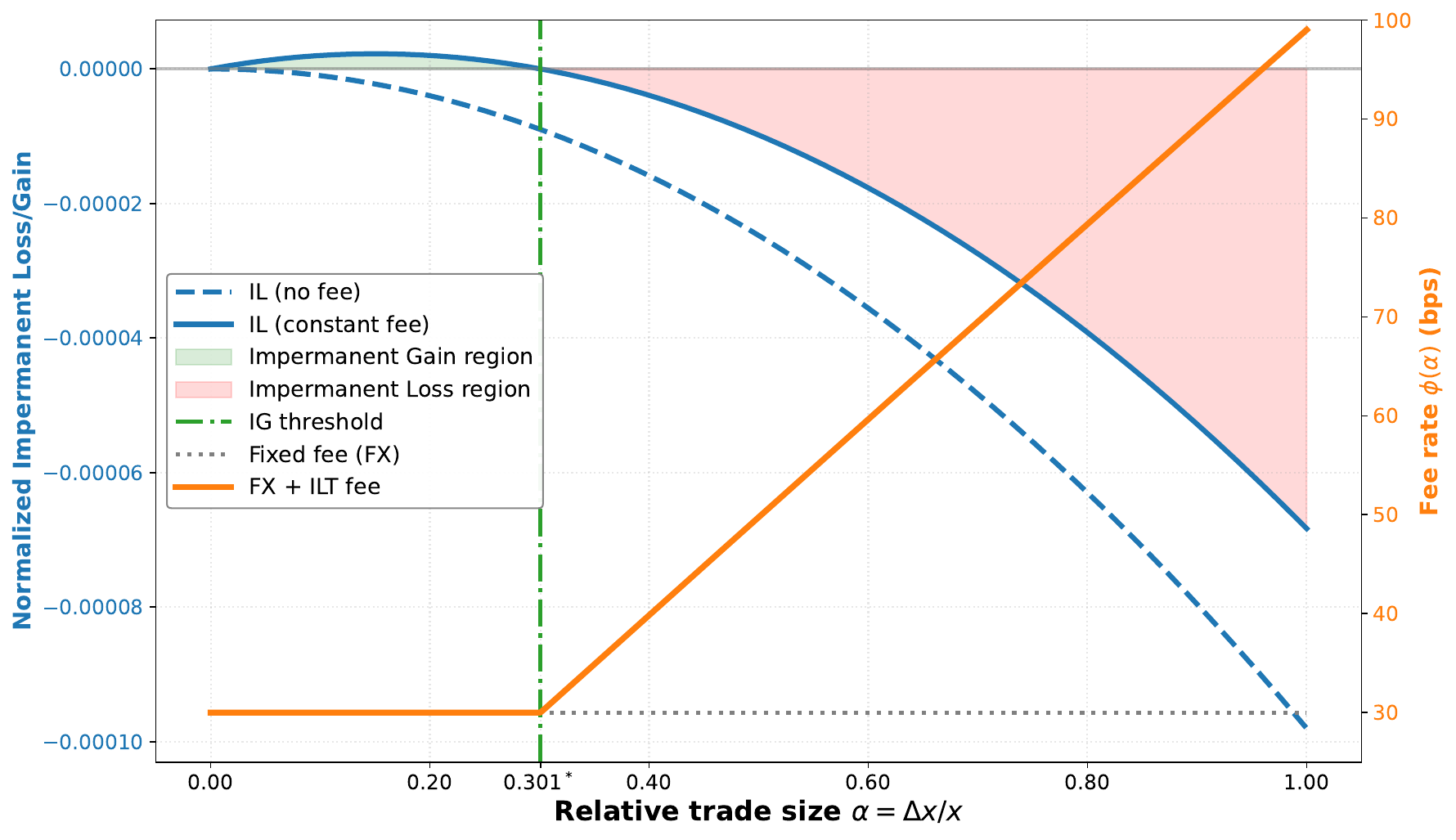}
    
    \caption{
        Visualization of the \textbf{Impermanent-Loss Trimming (ILT)} mechanism. 
        The left vertical axis represents the normalized Impermanent Loss/Gain (IL) for the liquidity provider, while the right vertical axis represents the fee rate $\phi(\alpha)$ in basis points (bps). 
        The solid blue curve shows IL under a constant fee, while the dashed blue curve shows IL without fees.
        The green shaded region corresponds to the \textbf{Impermanent Gain (IG)} zone where $\alpha \leq \alpha_{\max}$, and the red shaded region corresponds to the Impermanent Loss zone. 
        The orange curve illustrates the proposed fee structure: it remains equal to the fixed baseline fee (FX) within the IG region and scales linearly with trade size in the IL region to offset losses.
    }
    \label{fig:ilt_fee_illustration}
\end{figure}

This choice ensures that any trade exceeding the IG threshold pays exactly enough in fees to offset the resulting impermanent loss, effectively implementing a $[\text{IL}]_+$ penalty. Importantly, the modification is \emph{non-intrusive}: when $\alpha \leq \alpha_{\max}$, the fee remains at its baseline level, preserving low costs for benign trades.

\paragraph{Optimal arbitrage under ILT fees.}
For an Informed User maximizing $\Delta P(\alpha) = -p_A x \alpha + p_B y \cdot \frac{\alpha \gamma(\alpha)}{1+\alpha \gamma(\alpha)}$, the first-order condition becomes piecewise:
\begin{equation}
\frac{d}{d\alpha}\Delta P(\alpha) = 
\begin{cases}
-p_A x + p_B y \cdot \frac{\gamma_{\mathrm{fix}}}{(1+\alpha \gamma_{\mathrm{fix}})^2} = 0, & \alpha \leq \alpha_{\max} \\
-p_A x + p_B y \cdot \frac{1}{(1+2\alpha)^2} = 0, & \alpha > \alpha_{\max}
\end{cases}
\end{equation}
yielding optimal swap sizes:
\begin{equation}
\alpha^* = 
\begin{cases}
\frac{1}{\gamma_{\mathrm{fix}}}\left(\sqrt{\frac{p_B y \gamma_{\mathrm{fix}}}{p_A x}} - 1\right), & \text{if } \alpha^* \leq \alpha_{\max} \\
\frac{1}{2}\left(\sqrt{\frac{p_B y}{p_A x}} - 1\right), & \text{otherwise}
\end{cases}
\end{equation}
This demonstrates that ILT fees naturally dampen aggressive rebalancing by reducing the optimal arbitrage size for large price discrepancies, while preserving profitability for small deviations.

\section{Numerical Results}
\label{sec::results}

We evaluate the proposed framework using a liquidity pool of 25 million USDT split evenly between two tokens with IU and UU~\cite{lebedeva2025dynamic_fee} traders modeled as above. The analysis combines both synthetic price trajectories generated via Merton Jump-Diffusion (MJD) model~\cite{merton1976jump} and historical backtests across three different asset risk profiles introduced below. 

We categorize market conditions into four regimes -- Bull, Bear, Volatile and Calm -- based on the slope of the log-price regression and the standard deviation of hourly log-prices. This classification is applied to three categories of liquidity pools that represent varying risk profiles: stable-to-stable pairs~(e.g. USTC/USDT), stable-to-volatile asset pairs~(e.g., BTC/USDT), and volatile-to-volatile pairs~(e.g., ETH/BTC).

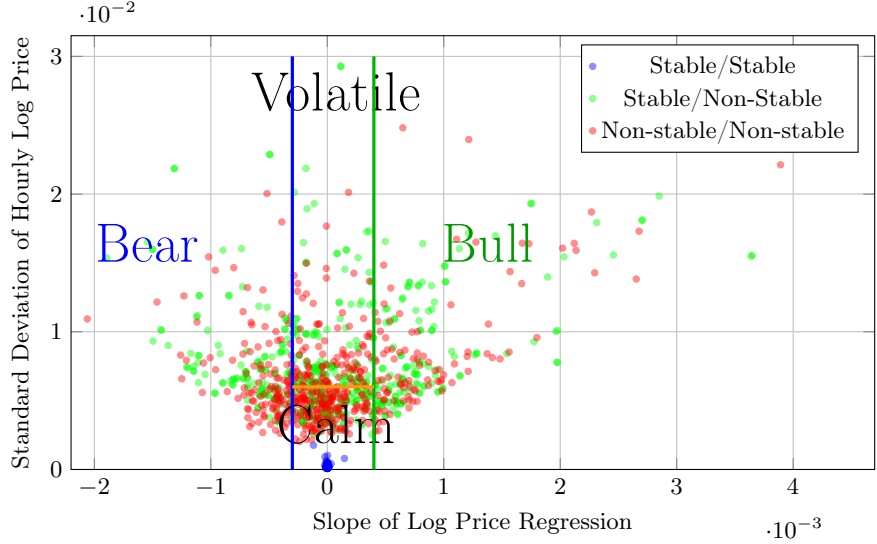
\begin{figure}[ht]
\centering
\begin{tikzpicture}
\begin{axis}[
    width=\textwidth,
    height=0.6\textwidth,
    xlabel={Slope of Log Price Regression},
    ylabel={Standard Deviation of Hourly Log Price},
    xmin=-0.0022,
    xmax=0.0047,
    ymin=0.0,
    ymax=0.0315,
    grid=major,
    clip=false,
    set layers=standard,
]

\addplot[
    only marks,
    mark=*,
    mark size=1.2pt,
    blue,
    opacity=0.45,
] table[
    col sep=comma,
    x=Slope,
    y=Std_dev
] {data/df_stable_stable.csv};
\addlegendentry{Stable/Stable}

\addplot[
    only marks,
    mark=*,
    mark size=1.2pt,
    green,
    opacity=0.45,
] table[
    col sep=comma,
    x=Slope,
    y=Std_dev
] {data/df_stable_not_stable.csv};
\addlegendentry{Stable/Non-Stable}

\addplot[
    only marks,
    mark=*,
    mark size=1.2pt,
    red,
    opacity=0.45,
] table[
    col sep=comma,
    x=Slope,
    y=Std_dev
] {data/df_not_stable_not_stable.csv};
\addlegendentry{Non-stable/Non-stable}

\addplot[
    blue,
    very thick,
    domain=0:0.03,
    samples=2,
    on layer=axis foreground,
] ({-0.00030},x);

\addplot[
    green!70!black,
    very thick,
    domain=0:0.03,
    samples=2,
    on layer=axis foreground,
] ({0.00040},x);

\addplot[
    orange!80!yellow,
    very thick,
    domain=-0.00030:0.00040,
    samples=2,
    on layer=axis foreground,
] (x,{0.00600});
\end{axis}
\node[font=\fontsize{20}{22}\selectfont, text=blue]
    at (1.0,3.0) {Bear};

\node[font=\fontsize{20}{22}\selectfont, text=green!60!black]
    at (5.5,3.0) {Bull};

\node[font=\fontsize{20}{22}\selectfont, text=black]
    at (3.5,5.0) {Volatile};

\node[font=\fontsize{20}{22}\selectfont, text=black]
    at (3.5,0.6) {Calm};
\end{tikzpicture}
\caption{Market segmentation into Bull, Bear, Volatile, and Calm regimes. A Bull regime is characterized by a positive price trend, a Bear regime by a negative price trend, a Volatile regime by large price fluctuations regardless of trend direction, and a Calm regime by low price variability and relatively stable market behavior.}
\label{fig:market_regimes}
\end{figure}

The classification, illustrated in Figure~\ref{fig:market_regimes}, partitions a one-year historical dataset into weekly segments. Bull and Bear regimes correspond to the top and bottom 25\% of regression slope values, identifying periods of sustained directional momentum. The remaining 50\% of the data, where no strong trend is present, is further split by the median standard deviation of log prices: intervals above the median are classified as Volatile, those below as Calm.

\subsection{Synthetic data}
To capture the fat-tailed return distributions and discrete price gaps that Geometric Brownian Motion (GBM) does not cover properly, we generate synthetic trajectories using Merton Jump-Diffusion (MJD) model~\cite{merton1976jump}: 
\begin{equation}
    \frac{d S_t}{S_{t-}} = (\mu - \lambda \kappa) dt + \sigma d W_t + d\left( \sum_{i = 1}^{N_t} \left(Y_i - 1\right) \right).
\end{equation}
Here $W_t$ is a standard Wiener process modeling market noise, $N_t$ is a Poisson process with intensity $\lambda$ generating discrete jumps, log-jumps magnitudes are normally distributed $\ln Y_t \sim \mathcal{N}(\mu_\text{jmp}, \sigma^2_\text{jmp})$, and $\kappa = \mathbb{E}\left[ Y_i - 1\right]$ is a drift compensator for the jump term. The drift $\mu$ and volatility $\sigma$ are calibrated from historical 12-second candle data, matching the approximate block time of Ethereum. Calibration differs by asset class: for volatile pairs we use larger $\lambda$ and $\sigma_\text{jmp}$ to reproduce flash-event behavior, while for stablecoins jumps are suppressed $(\sigma_\text{jmp} \rightarrow 0)$ to reflect their pegged nature. We run Monte Carlo simulations across $1000$ random seeds and compute the accumulated LP yield under each fee algorithm.

Tables \ref{tab:usdc_rad_bear}--\ref{tab:doge_btc_volatile} report LP, IU and UU yields across the four market regimes. A consistent pattern emerges: ILT-enhanced configurations improve LP outcomes across all scenarios while reducing IU yields. The endogenous AMM specification, which instantiates the AMM-in-AMM architecture, dominates the reactive baselines in trending markets (Tables \ref{tab:usdc_rad_bear}, \ref{tab:usdc_rad_bull}), reaching $27.50 \; \text{bps}$ and $26.99 \; \text{bps}$ in bear and bull regimes under ILT -- improvements of roughly 6–7\% over the AMM baseline. The most pronounced gains appear in the calm and volatile regimes (Tables \ref{tab:eth_btc_calm}, \ref{tab:doge_btc_volatile}): in the calm regime, the ILT-enhanced AMM more than doubles its baseline LP yield ($10.21 \rightarrow 22.37 \; \text{bps}$), and under volatile conditions ILT increases LP yield by approximately 24\% (to $20.58 \; \text{bps}$).

Across all synthetic evaluations, the UU yield remains stable at $-57.2 \pm 1.2 \; \text{bps}$ (standard deviation below 2\%), confirming that ILT is non-intrusive for small, utility-driven trades. IU yield decreases by $3.2 \pm 1.8\%$ on average under ILT, indicating that the mechanism captures value from large, information-driven executions that exceed the IG threshold $\alpha_\text{max}$.

\newcolumntype{C}{>{\centering\arraybackslash}X}

\begin{table}[!htbp]
\centering
\scriptsize
\setlength{\tabcolsep}{2.5pt}
\renewcommand{\arraystretch}{0.8}
\caption{Synthetic data yields across asset pairs and market regimes.}
\label{tab:synthetic_results_all}

\begin{subtable}{\textwidth}
\centering
\caption{USDC/RAD (Bear Market)}
\label{tab:usdc_rad_bear}
\begin{tabularx}{\linewidth}{@{}CCCC@{}}
\toprule
Alg & IU & UU & LP \\
\midrule
AMM      & $9.56 \pm 2.67$ & $-57.25 \pm 1.13$ & $25.78 \pm 2.61$ \\
ILT\_AMM & $9.20 \pm 2.39$ & $-57.22 \pm 1.15$ & $27.50 \pm 1.74$ \\
\midrule
BA       & $8.94 \pm 2.30$ & $-57.24 \pm 1.16$ & $20.96 \pm 2.32$ \\
ILT\_BA  & $8.76 \pm 2.11$ & $-57.25 \pm 1.17$ & $22.15 \pm 1.76$ \\
\midrule
DA       & $8.72 \pm 2.16$ & $-57.53 \pm 1.16$ & $19.14 \pm 2.48$ \\
ILT\_DA  & $8.58 \pm 1.97$ & $-57.50 \pm 1.14$ & $20.26 \pm 2.00$ \\
\midrule
OB       & $9.20 \pm 2.62$ & $-57.26 \pm 0.98$ & $30.63 \pm 1.78$ \\
ILT\_OB  & $9.04 \pm 2.43$ & $-57.27 \pm 0.98$ & $31.73 \pm 0.92$ \\
\midrule
FX       & $8.78 \pm 2.19$ & $-57.19 \pm 1.21$ & $19.59 \pm 1.66$ \\
ILT\_FX  & $8.62 \pm 2.01$ & $-57.17 \pm 1.19$ & $20.73 \pm 0.83$ \\
\bottomrule
\end{tabularx}
\end{subtable}

\vspace{0.15em}

\begin{subtable}{\textwidth}
\centering
\caption{USDC/RAD (Bull Market)}
\label{tab:usdc_rad_bull}
\begin{tabularx}{\linewidth}{@{}CCCC@{}}
\toprule
Alg & IU & UU & LP \\
\midrule
AMM      & $9.37 \pm 1.43$ & $-56.37 \pm 1.13$ & $25.89 \pm 1.82$ \\
ILT\_AMM & $9.16 \pm 1.24$ & $-56.31 \pm 1.20$ & $26.99 \pm 1.31$ \\
\midrule
BA       & $8.99 \pm 1.21$ & $-56.40 \pm 1.16$ & $20.08 \pm 1.56$ \\
ILT\_BA  & $8.91 \pm 1.12$ & $-56.39 \pm 1.16$ & $20.78 \pm 1.21$ \\
\midrule
DA       & $8.96 \pm 1.15$ & $-56.41 \pm 1.34$ & $18.56 \pm 1.75$ \\
ILT\_DA  & $8.89 \pm 1.07$ & $-56.42 \pm 1.33$ & $19.20 \pm 1.49$ \\
\midrule
OB       & $9.13 \pm 1.40$ & $-56.46 \pm 1.20$ & $31.41 \pm 1.11$ \\
ILT\_OB  & $9.05 \pm 1.29$ & $-56.46 \pm 1.20$ & $32.08 \pm 0.59$ \\
\midrule
FX       & $8.96 \pm 1.16$ & $-56.27 \pm 1.26$ & $18.75 \pm 0.98$ \\
ILT\_FX  & $8.88 \pm 1.07$ & $-56.29 \pm 1.27$ & $19.45 \pm 0.50$ \\
\bottomrule
\end{tabularx}
\end{subtable}

\vspace{0.15em}

\begin{subtable}{\textwidth}
\centering
\caption{ETH/BTC (Calm Market)}
\label{tab:eth_btc_calm}
\begin{tabularx}{\linewidth}{@{}CCCC@{}}
\toprule
Alg & IU & UU & LP \\
\midrule
AMM      & $45.66 \pm 17.84$ & $-56.74 \pm 0.93$ & $10.21 \pm 9.24$ \\
ILT\_AMM & $41.08 \pm 15.06$ & $-56.72 \pm 0.94$ & $22.37 \pm 2.67$ \\
\midrule
BA       & $43.49 \pm 17.32$ & $-56.68 \pm 1.01$ & $9.25 \pm 9.11$ \\
ILT\_BA  & $39.73 \pm 14.51$ & $-56.74 \pm 0.99$ & $19.18 \pm 3.46$ \\
\midrule
DA       & $43.98 \pm 17.22$ & $-56.73 \pm 1.02$ & $8.27 \pm 9.34$ \\
ILT\_DA  & $40.00 \pm 14.00$ & $-56.77 \pm 0.99$ & $18.46 \pm 3.33$ \\
\midrule
OB       & $45.86 \pm 18.79$ & $-56.68 \pm 0.92$ & $15.66 \pm 8.34$ \\
ILT\_OB  & $42.19 \pm 15.85$ & $-56.71 \pm 0.92$ & $24.58 \pm 2.40$ \\
\midrule
FX       & $43.94 \pm 17.26$ & $-56.70 \pm 0.99$ & $8.68 \pm 9.07$ \\
ILT\_FX  & $39.94 \pm 14.02$ & $-56.74 \pm 0.94$ & $18.84 \pm 2.97$ \\
\bottomrule
\end{tabularx}
\end{subtable}

\vspace{0.15em}

\begin{subtable}{\textwidth}
\centering
\caption{DOGE/BTC (Volatile Market)}
\label{tab:doge_btc_volatile}
\begin{tabularx}{\linewidth}{@{}CCCC@{}}
\toprule
Alg & IU & UU & LP \\
\midrule
AMM      & $18.89 \pm 3.76$ & $-55.33 \pm 1.48$ & $16.60 \pm 4.18$ \\
ILT\_AMM & $17.81 \pm 3.03$ & $-55.32 \pm 1.44$ & $20.58 \pm 2.57$ \\
\midrule
BA       & $19.72 \pm 3.50$ & $-55.50 \pm 1.36$ & $12.93 \pm 4.90$ \\
ILT\_BA  & $19.13 \pm 3.08$ & $-55.52 \pm 1.37$ & $15.84 \pm 3.80$ \\
\midrule
DA       & $17.28 \pm 3.55$ & $-55.44 \pm 1.39$ & $12.28 \pm 3.75$ \\
ILT\_DA  & $16.72 \pm 3.08$ & $-55.46 \pm 1.41$ & $15.03 \pm 2.55$ \\
\midrule
OB       & $12.42 \pm 3.34$ & $-55.55 \pm 1.22$ & $28.80 \pm 2.67$ \\
ILT\_OB  & $11.90 \pm 2.89$ & $-55.56 \pm 1.23$ & $31.37 \pm 0.89$ \\
\midrule
FX       & $15.32 \pm 3.56$ & $-55.40 \pm 1.43$ & $14.26 \pm 2.91$ \\
ILT\_FX  & $14.74 \pm 3.06$ & $-55.40 \pm 1.42$ & $16.98 \pm 1.23$ \\
\bottomrule
\end{tabularx}
\end{subtable}

\end{table}

\subsection{Historical data}
We employ the Dolan-Mor\'e performance profile methodology~\cite{dolan2002benchmarking}, which compares algorithms by the fraction of problems on which each performs within a factor $\tau$ of the best result. Defining a problem $p \in P$ as a unique combination of a trading pair and a specific market regime, a fee algorithm $s \in S$ as a solver, and the LP markout $x_{p, s}$ as a performance metric to be maximized, the performance ratio is defined as
\begin{equation}
r_{p,s} = 
\begin{cases}
    \frac{\max_{s' \in S} x_{p,s'}}{x_{p,s}}, & x_{p, s} > 0, \\
    \infty, & x_{p,s} \leq 0. 
\end{cases}
\end{equation}

The performance profile of algorithm $s$ is the cumulative distribution function: 
\begin{equation}
    \rho_s(\tau) = \frac{1}{|P|} \cdot \text{size}\left\{ p \in P : r_{p, s} \leq \tau \right\}, 
\end{equation}
giving the fraction of problems on which $s$ lies within factor $\tau$ of the best algorithm.

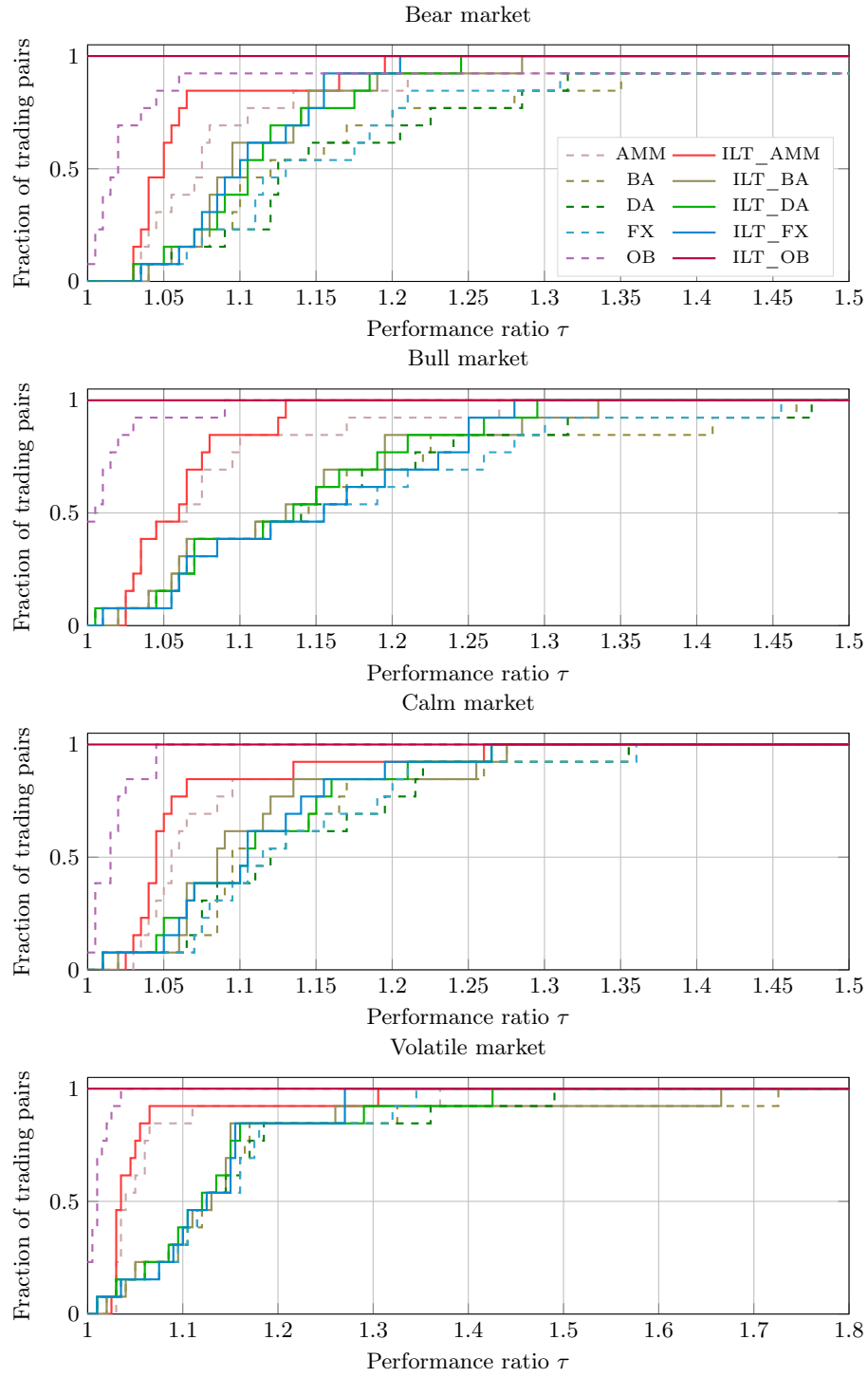
\begin{figure}[!htbp]
\centering
\begin{tikzpicture}
\begin{groupplot}[
    group style={
        group size=1 by 4,
        vertical sep=1.5cm
    },
    width=\textwidth,
    height=0.4\textwidth,
    ymin=0,
    ymax=1.05,
    grid=major,
    xlabel={Performance ratio $\tau$},
    ylabel={Fraction of trading pairs},
    tick label style={font=\small},
    label style={font=\small},
    title style={font=\small},
    legend style={
        at={(0.98,0.02)},
        anchor=south east,
        font=\scriptsize,
        draw=gray!30,
        fill=white,
        legend columns=2
    }
]

\nextgroupplot[
    title={Bear market},
    xmin=1,
    xmax=1.5
]

\addplot[pink!60!gray, dashed, thick, const plot]
table[col sep=comma, x=tau, y=rho_AMM_fee, each nth point=5] {data/perfprof_bear_market.csv};
\addlegendentry{AMM}

\addplot[red!70!pink, thick, const plot]
table[col sep=comma, x=tau, y=rho_IMPROVED_AMM_fee, each nth point=5] {data/perfprof_bear_market.csv};
\addlegendentry{ILT\_AMM}

\addplot[olive!70!gray, dashed, thick, const plot]
table[col sep=comma, x=tau, y=rho_BA_fee, each nth point=5] {data/perfprof_bear_market.csv};
\addlegendentry{BA}

\addplot[yellow!50!black, thick, const plot]
table[col sep=comma, x=tau, y=rho_IMPROVED_BA_fee, each nth point=5] {data/perfprof_bear_market.csv};
\addlegendentry{ILT\_BA}

\addplot[green!50!black, dashed, thick, const plot]
table[col sep=comma, x=tau, y=rho_DA_fee, each nth point=5] {data/perfprof_bear_market.csv};
\addlegendentry{DA}

\addplot[green!70!black, thick, const plot]
table[col sep=comma, x=tau, y=rho_IMPROVED_DA_fee, each nth point=5] {data/perfprof_bear_market.csv};
\addlegendentry{ILT\_DA}

\addplot[cyan!60!gray, dashed, thick, const plot]
table[col sep=comma, x=tau, y=rho_FX_fee, each nth point=5] {data/perfprof_bear_market.csv};
\addlegendentry{FX}

\addplot[cyan!70!blue, thick, const plot]
table[col sep=comma, x=tau, y=rho_IMPROVED_FX_fee, each nth point=5] {data/perfprof_bear_market.csv};
\addlegendentry{ILT\_FX}

\addplot[violet!60, dashed, thick, const plot]
table[col sep=comma, x=tau, y=rho_OB_fee, each nth point=5] {data/perfprof_bear_market.csv};
\addlegendentry{OB}

\addplot[purple, thick, const plot]
table[col sep=comma, x=tau, y=rho_IMPROVED_OB_fee, each nth point=5] {data/perfprof_bear_market.csv};
\addlegendentry{ILT\_OB}

\nextgroupplot[
    title={Bull market},
    xmin=1,
    xmax=1.5
]

\addplot[pink!60!gray, dashed, thick, const plot]
table[col sep=comma, x=tau, y=rho_AMM_fee, each nth point=5] {data/perfprof_bull_market.csv};

\addplot[red!70!pink, thick, const plot]
table[col sep=comma, x=tau, y=rho_IMPROVED_AMM_fee, each nth point=5] {data/perfprof_bull_market.csv};

\addplot[olive!70!gray, dashed, thick, const plot]
table[col sep=comma, x=tau, y=rho_BA_fee, each nth point=5] {data/perfprof_bull_market.csv};

\addplot[yellow!50!black, thick, const plot]
table[col sep=comma, x=tau, y=rho_IMPROVED_BA_fee, each nth point=5] {data/perfprof_bull_market.csv};

\addplot[green!50!black, dashed, thick, const plot]
table[col sep=comma, x=tau, y=rho_DA_fee, each nth point=5] {data/perfprof_bull_market.csv};

\addplot[green!70!black, thick, const plot]
table[col sep=comma, x=tau, y=rho_IMPROVED_DA_fee, each nth point=5] {data/perfprof_bull_market.csv};

\addplot[cyan!60!gray, dashed, thick, const plot]
table[col sep=comma, x=tau, y=rho_FX_fee, each nth point=5] {data/perfprof_bull_market.csv};

\addplot[cyan!70!blue, thick, const plot]
table[col sep=comma, x=tau, y=rho_IMPROVED_FX_fee, each nth point=5] {data/perfprof_bull_market.csv};

\addplot[violet!60, dashed, thick, const plot]
table[col sep=comma, x=tau, y=rho_OB_fee, each nth point=5] {data/perfprof_bull_market.csv};

\addplot[purple, thick, const plot]
table[col sep=comma, x=tau, y=rho_IMPROVED_OB_fee, each nth point=5] {data/perfprof_bull_market.csv};

\nextgroupplot[
    title={Calm market},
    xmin=1,
    xmax=1.5
]

\addplot[pink!60!gray, dashed, thick, const plot]
table[col sep=comma, x=tau, y=rho_AMM_fee, each nth point=5] {data/perfprof_calm_market.csv};

\addplot[red!70!pink, thick, const plot]
table[col sep=comma, x=tau, y=rho_IMPROVED_AMM_fee, each nth point=5] {data/perfprof_calm_market.csv};

\addplot[olive!70!gray, dashed, thick, const plot]
table[col sep=comma, x=tau, y=rho_BA_fee, each nth point=5] {data/perfprof_calm_market.csv};

\addplot[yellow!50!black, thick, const plot]
table[col sep=comma, x=tau, y=rho_IMPROVED_BA_fee, each nth point=5] {data/perfprof_calm_market.csv};

\addplot[green!50!black, dashed, thick, const plot]
table[col sep=comma, x=tau, y=rho_DA_fee, each nth point=5] {data/perfprof_calm_market.csv};

\addplot[green!70!black, thick, const plot]
table[col sep=comma, x=tau, y=rho_IMPROVED_DA_fee, each nth point=5] {data/perfprof_calm_market.csv};

\addplot[cyan!60!gray, dashed, thick, const plot]
table[col sep=comma, x=tau, y=rho_FX_fee, each nth point=5] {data/perfprof_calm_market.csv};

\addplot[cyan!70!blue, thick, const plot]
table[col sep=comma, x=tau, y=rho_IMPROVED_FX_fee, each nth point=5] {data/perfprof_calm_market.csv};

\addplot[violet!60, dashed, thick, const plot]
table[col sep=comma, x=tau, y=rho_OB_fee, each nth point=5] {data/perfprof_calm_market.csv};

\addplot[purple, thick, const plot]
table[col sep=comma, x=tau, y=rho_IMPROVED_OB_fee, each nth point=5] {data/perfprof_calm_market.csv};

\nextgroupplot[
    title={Volatile market},
    xmin=1,
    xmax=1.8
]

\addplot[pink!60!gray, dashed, thick, const plot]
table[col sep=comma, x=tau, y=rho_AMM_fee, each nth point=5] {data/perfprof_volatile_market.csv};

\addplot[red!70!pink, thick, const plot]
table[col sep=comma, x=tau, y=rho_IMPROVED_AMM_fee, each nth point=5] {data/perfprof_volatile_market.csv};

\addplot[olive!70!gray, dashed, thick, const plot]
table[col sep=comma, x=tau, y=rho_BA_fee, each nth point=5] {data/perfprof_volatile_market.csv};

\addplot[yellow!50!black, thick, const plot]
table[col sep=comma, x=tau, y=rho_IMPROVED_BA_fee, each nth point=5] {data/perfprof_volatile_market.csv};

\addplot[green!50!black, dashed, thick, const plot]
table[col sep=comma, x=tau, y=rho_DA_fee, each nth point=5] {data/perfprof_volatile_market.csv};

\addplot[green!70!black, thick, const plot]
table[col sep=comma, x=tau, y=rho_IMPROVED_DA_fee, each nth point=5] {data/perfprof_volatile_market.csv};

\addplot[cyan!60!gray, dashed, thick, const plot]
table[col sep=comma, x=tau, y=rho_FX_fee, each nth point=5] {data/perfprof_volatile_market.csv};

\addplot[cyan!70!blue, thick, const plot]
table[col sep=comma, x=tau, y=rho_IMPROVED_FX_fee, each nth point=5] {data/perfprof_volatile_market.csv};

\addplot[violet!60, dashed, thick, const plot]
table[col sep=comma, x=tau, y=rho_OB_fee, each nth point=5] {data/perfprof_volatile_market.csv};

\addplot[purple, thick, const plot]
table[col sep=comma, x=tau, y=rho_IMPROVED_OB_fee, each nth point=5] {data/perfprof_volatile_market.csv};

\end{groupplot}
\end{tikzpicture}
\caption{Dolan--Mor\'e performance profiles for different market regimes. Dashed lines denote the original methods, and solid lines denote the improved methods.}
\label{fig:performance_profiles}
\end{figure}

Figure~\ref{fig:performance_profiles} reports the Dolan–Mor\'e profiles across the four market regimes. ILT-enhanced methods (solid lines) systematically dominate their non-ILT counterparts (dashed lines) in every regime. The effect is most pronounced for the AMM-in-AMM specification, whose ILT-enhanced curve rises substantially faster and reaches a high fraction of trading pairs already at low $\tau$, indicating that the trimming mechanism makes this method robust across a broad range of pairs. Improvements for the BA, DA, and FX specifications are smaller but consistent, with the largest separation visible in bear and bull regimes.

Among the original methods (dashed curves), the oracle-based benchmark (OB) is the strongest performer across all regimes, with the AMM specification ranking next among implementable algorithms. BA, DA, and FX form a middle-performance group whose internal ordering depends on the regime. Under the volatile regime, differences between methods compress, but ILT-enhanced versions retain at least weak dominance over their baselines -- supporting the interpretation that the trimming contribution is stable rather than regime-specific.

Aggregating across all 36 historical scenarios (9 pairs × 4 regimes), ILT-enhanced algorithms achieve a median LP yield improvement of 18.3\% over their baselines, with the largest gains in volatile non-stable/non-stable pairs (e.g., DOGE/BTC: +16.6\% for FX, +24.0\% for AMM). The performance profile curves (Figure~\ref{fig:performance_profiles}) confirm that ILT variants reach $\rho(\tau=1.05) > 0.65$ in all regimes, meaning they perform within 5\% of the best algorithm on at least 65\% of scenarios.

\begin{figure}[!htbp]
\centering
\begin{tikzpicture}
\begin{groupplot}[
    group style={
        group size=1 by 2,
        vertical sep=1.5cm
    },
    width=\textwidth,
    height=0.4\textwidth,
    ymin=0,
    ymax=1.05,
    grid=major,
    xlabel={Performance ratio $\tau$},
    ylabel={Fraction of trading pairs},
    tick label style={font=\small},
    label style={font=\small},
    title style={font=\small},
    legend style={
        at={(0.98,0.02)},
        anchor=south east,
        font=\scriptsize,
        draw=gray!30,
        fill=white,
        legend columns=2
    }
]

\nextgroupplot[
    title={IU},
    xmin=1,
    xmax=1.5
]

\addplot[pink!60!gray, dashed, thick, const plot]
table[col sep=comma, x=tau, y=rho_AMM_fee, each nth point=5] {data/informed_aggregated_performance.csv};
\addlegendentry{AMM}

\addplot[red!70!pink, thick, const plot]
table[col sep=comma, x=tau, y=rho_IMPROVED_AMM_fee, each nth point=5] {data/informed_aggregated_performance.csv};
\addlegendentry{ILT\_AMM}

\addplot[olive!70!gray, dashed, thick, const plot]
table[col sep=comma, x=tau, y=rho_BA_fee, each nth point=5] {data/informed_aggregated_performance.csv};
\addlegendentry{BA}

\addplot[yellow!50!black, thick, const plot]
table[col sep=comma, x=tau, y=rho_IMPROVED_BA_fee, each nth point=5] {data/informed_aggregated_performance.csv};
\addlegendentry{ILT\_BA}

\addplot[green!50!black, dashed, thick, const plot]
table[col sep=comma, x=tau, y=rho_DA_fee, each nth point=5] {data/informed_aggregated_performance.csv};
\addlegendentry{DA}

\addplot[green!70!black, thick, const plot]
table[col sep=comma, x=tau, y=rho_IMPROVED_DA_fee, each nth point=5] {data/informed_aggregated_performance.csv};
\addlegendentry{ILT\_DA}

\addplot[cyan!60!gray, dashed, thick, const plot]
table[col sep=comma, x=tau, y=rho_FX_fee, each nth point=5] {data/informed_aggregated_performance.csv};
\addlegendentry{FX}

\addplot[cyan!70!blue, thick, const plot]
table[col sep=comma, x=tau, y=rho_IMPROVED_FX_fee, each nth point=5] {data/informed_aggregated_performance.csv};
\addlegendentry{ILT\_FX}

\addplot[violet!60, dashed, thick, const plot]
table[col sep=comma, x=tau, y=rho_OB_fee, each nth point=5] {data/informed_aggregated_performance.csv};
\addlegendentry{OB}

\addplot[purple, thick, const plot]
table[col sep=comma, x=tau, y=rho_IMPROVED_OB_fee, each nth point=5] {data/informed_aggregated_performance.csv};
\addlegendentry{ILT\_OB}

\nextgroupplot[
    title={UU},
    xmin=1,
    xmax=1.01
]

\addplot[pink!60!gray, dashed, thick, const plot]
table[col sep=comma, x=tau, y=rho_AMM_fee, each nth point=5] {data/uninformed_aggregated_performance.csv};

\addplot[red!70!pink, thick, const plot]
table[col sep=comma, x=tau, y=rho_IMPROVED_AMM_fee, each nth point=5] {data/uninformed_aggregated_performance.csv};

\addplot[olive!70!gray, dashed, thick, const plot]
table[col sep=comma, x=tau, y=rho_BA_fee, each nth point=5] {data/uninformed_aggregated_performance.csv};

\addplot[yellow!50!black, thick, const plot]
table[col sep=comma, x=tau, y=rho_IMPROVED_BA_fee, each nth point=5] {data/uninformed_aggregated_performance.csv};

\addplot[green!50!black, dashed, thick, const plot]
table[col sep=comma, x=tau, y=rho_DA_fee, each nth point=5] {data/uninformed_aggregated_performance.csv};

\addplot[green!70!black, thick, const plot]
table[col sep=comma, x=tau, y=rho_IMPROVED_DA_fee, each nth point=5] {data/uninformed_aggregated_performance.csv};

\addplot[cyan!60!gray, dashed, thick, const plot]
table[col sep=comma, x=tau, y=rho_FX_fee, each nth point=5] {data/uninformed_aggregated_performance.csv};

\addplot[cyan!70!blue, thick, const plot]
table[col sep=comma, x=tau, y=rho_IMPROVED_FX_fee, each nth point=5] {data/uninformed_aggregated_performance.csv};

\addplot[violet!60, dashed, thick, const plot]
table[col sep=comma, x=tau, y=rho_OB_fee, each nth point=5] {data/uninformed_aggregated_performance.csv};

\addplot[purple, thick, const plot]
table[col sep=comma, x=tau, y=rho_IMPROVED_OB_fee, each nth point=5] {data/uninformed_aggregated_performance.csv};

\end{groupplot}
\end{tikzpicture}
\caption{Dolan--Mor\'e performance profiles for the IU and UU settings. Dashed lines denote the original methods, and solid lines denote the improved methods.}
\label{fig:iu_uu_performance_profiles}
\end{figure}
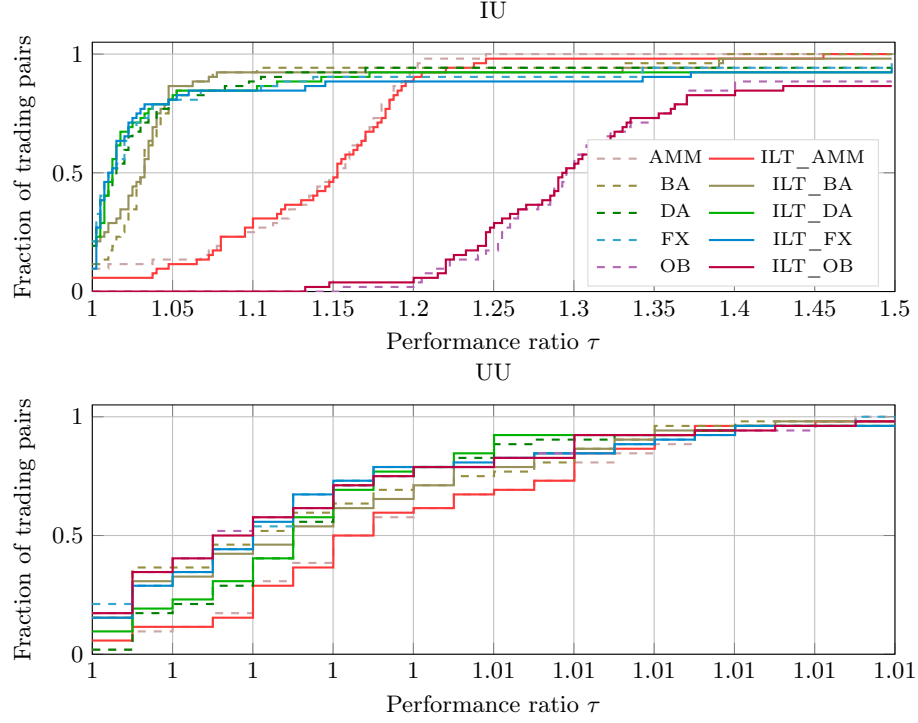

Figure \ref{fig:iu_uu_performance_profiles} reports the aggregated Dolan-Mor\'e performance profiles for IU (top panel) and UU (bottom panel). The IU profiles show that fee specifications more closely aligned with LP protection shift IU curves to the right -- informed traders attain near-best outcomes for a smaller fraction of scenarios at any given $\tau$, with the effect most visible for the AMM-in-AMM and oracle-based specifications. The UU profiles, in contrast, are tightly clustered and reach near-unity by $\tau = 1.01$, confirming that the proposed mechanisms do not materially alter outcomes for uninformed participants. Combined with the UU yield stability documented above, this indicates that the proposed mechanisms reallocate value from informed arbitrageurs to LPs without imposing meaningful costs on benign flow.

\section{Conclusion}
\label{sec:conclusion}

We have shown that dynamic fee design in AMMs should be treated as a mechanism-design problem rather than a reactive tuning task. The proposed AMM-in-AMM architecture endogenizes fees as state variables of a secondary invariant, enabling liquidity and fee states to co-evolve. The Impermanent-Loss Trimming (ILT) modification then provides a principled, trade-size-aware criterion for distinguishing benign flow from risk-intensive arbitrage.

Empirically, ILT-enhanced algorithms improve LP yields by 6--24\% in volatile markets and up to 119\% in calm regimes across synthetic and historical data, while preserving uninformed user participation (UU yield stable at $-57.2 \pm 1.2$ bps) and reducing informed arbitrage profitability by $3.2 \pm 1.8\%$. Performance profiles confirm ILT variants dominate baselines across 60--75\% of scenarios.

These results imply: (1) trade-size-aware fees provide a practical foundation for programmable DEXs (e.g., via Uniswap v4 hooks); (2) endogenous fee mechanisms better align transaction costs with LP risk; and (3) structural fee adjustments can curb informed arbitrage without deterring benign participation.

Future work should address implementation frictions (gas costs, oracle latency), estimate the psychological loss factor $r$ from data, and extend the framework to concentrated-liquidity AMMs.

A natural concern is that informed users may evade ILT by splitting a large trade into smaller transactions within the IG region. By construction, every fragment then lies in the IG zone with non-negative per-trade impermanent loss -- the LP is shielded from large adverse rebalancing on any single transaction, regardless of the arbitrageur's strategy. Splitting reduces fee capture but is constrained by gas costs that scale linearly with the number of transactions; path-dependent ILT extensions that aggregate trade activity over short windows present a promising direction and are left to future work.

Overall, endogenous, trade-aware dynamic fees offer a structurally grounded path toward more sustainable and equitable liquidity provision in DeFi.

\bibliographystyle{splncs04_custom}
\bibliography{refs2025}

\end{document}